\documentclass[]{piparticle-final}
\usepackage{graphicx}
\usepackage{amsmath}

\usepackage{cite} 
\usepackage[normalem]{ulem}  
\usepackage{color} 
\usepackage{epstopdf}

\begin{document}
\volume{8}               
\articlenumber{080003}   
\journalyear{2016}       
\editor{L. A. Pugnaloni}   
\reviewers{C. M. Carlevaro, Instituto de F\'isica de L\'iquidos y Sistemas Biol\'ogicos,\\ \mbox{} \hspace{4.25cm} La Plata, Argentina.}  
\received{{22} November 2015}     
\accepted{19 February 2016}   
\runningauthor{A. Murray \itshape{et al.}}  
\doi{080003}         

\title{Increasing granular flow rate with obstructions}

\author{Alan Murray,\cite{inst1} 
        Fernando Alonso-Marroquin\cite{inst2}\thanks{E-mail: fernando.alonso@sydney.edu.au} }
        
\pipabstract{
We describe a simple experiment involving spheres rolling down an
inclined plane towards a bottleneck and through a gap. Results of
the experiment indicate that flow rate can be increased by placing
an obstruction at optimal positions near the bottleneck. We use the
experiment to develop a computer simulation using the PhysX physics
engine. Simulations confirm the experimental results and we state
several considerations necessary to obtain a model that agrees well
with experiment. We demonstrate that the model exhibits clogging,
intermittent and continuous flow, and that it can be used as a tool
for further investigations in granular flow. 
}

\maketitle

\blfootnote{
\begin{theaffiliation}{99}
   \institution{inst1} SAE Creative Media Institute, Sydney, Australia.
   \institution{inst2} School of Civil Engineering, The University of Sydney, NSW 2006, Australia.
\end{theaffiliation}
}

\section{Introduction}
\label{intro}

When does an obstruction placed near a bottleneck increase the flow rate
of discrete objects moving through the bottleneck? Answering this
question is of great utility on many scales. For example, the safe
evacuation of pedestrians moving through confined environments (train
stations, stadiums, concert halls, etc) in emergency situations can
be of vital importance. And grain falling through a silo, where efficient
flow is necessary for production and clogging is undesirable, is another
example.

As stated by Magalhaes et al. in Ref. \cite{Magalhaes}, `Granular materials
are ubiquitous either in nature or in industrial processes' and so
a fundamental understanding of their motions is of intrinsic interest,
both from physics and engineering perspectives. The quantitative study
of the flow of granular materials has been performed for many decades
\cite{Beverloo}, yet there is enormous scope for many interesting
and creative techniques to be developed: experimental, theoretical and computational. Granular flow around an obstruction placed near a
bottleneck is of particular interest as it has many potential applications
and this has been investigated by many authors \cite{Magalhaes,Wilson,Mort,Rodolfo,Lumay,Zuriguel2014}.

Increases in flow rate achieved through placement of an obstruction
near a bottleneck have been reported by several investigators \cite{Alonso-Marroquin,Yang,Zuriguel2011}.
Another commonly reported phenomenon is that of `clogging', also
described as the formation of arches \cite{Zuriguel2014}. There are
also many novel investigations of granular flow. For example, Wilson
et al. \cite{Wilson} consider granular flow of particles that are
completely submerged under water and Lumay et al. \cite{Lumay} investigate
the flow of charged particles that repel each other. Clearly, granular
flow is a very wide and active area of research.

Several authors have investigated the use of physics engines used
in games, an example of which can be found in the work by Carlevaro and Pugnaloni
\cite{Carlevaro}. In this paper we present a real time, 3D computer
simulation using the PhysX physics engine. The model can be used in
the study of granular flow as it exhibits continuous flow, intermittent
flow, clogging, and has several parameters that an investigator can
control as the simulation is occurring. The computer model is based
on an experiment that uses spheres for the particles and three different
obstruction shapes. We use experimental data to calibrate and validate
our model. The model can then be used to determine how the highest
flow rates may be achieved both with regard to obstruction position
and to find optimal obstruction shapes.

The paper is organized as follows: we present the model in section
\ref{sec:The Model} Then we describe the experiments in two parts.
Section III deals with flow rate without obstructions,
and section IV investigates flow rate with
obstructions. We discuss the limitations in section V,
and conclude in section \ref{sec:Conclusion}

\section{The Model}

\label{sec:The Model}

\textcolor{black}{We use PhysX \cite{PhysX}, a freely available real
time, 3D physics engine for computations. PhysX is widely used in
the games industry as, when necessary, it favors speed and stability
of computation over accuracy. Integration of the equations of motion
is done using an unconditionally stable, semi implicit Euler scheme
yielding algebraic equations that are solved using a progressive Gauss-Seidel
algorithm, whilst enforcing the Signorini conditions \cite{Tonge}.
The time step can be adjusted for greater accuracy at the cost of
speed and hence, real time performance of the engine. For this study,
we left the time step at its default value of $0.02$ seconds.}

\textcolor{black}{PhysX uses the Coulomb model for friction and restitution
is a measure of the loss of kinetic energy between colliding particles.
Details and further references about the rigid body system and friction
model used in PhysX can be found in the work by Tonge \cite{Tonge}. }

\textcolor{black}{The model incorporates the following parameters
which are freely specifiable and chosen to agree with experimental
results: friction, restitution, and sphere diameter. In addition,
our model allows us to vary the angle of inclination of the plane,
the gap width and the distance from the obstruction to the gap. We
incorporate imperfections in the sphere diameters by allowing them
to vary by up to 5\%. The model uses $1551$ particles whilst $\approx1000$
particles were used in the experiment. Each simulation is performed
$20$ times.}

\textcolor{black}{Unity \cite{Unity}, a freely available game engine,
is used as our programming interface to PhysX. We also use Blender
\cite{Blender}, a freely available 3D modeling package for modeling
the obstructions.}

\section{Flow rate without obstructions}

\label{sec:No Obstructions}

\begin{figure}[htb]
\centering
\includegraphics[width=0.95\columnwidth,keepaspectratio=true,trim=0 0 7cm 0,clip]{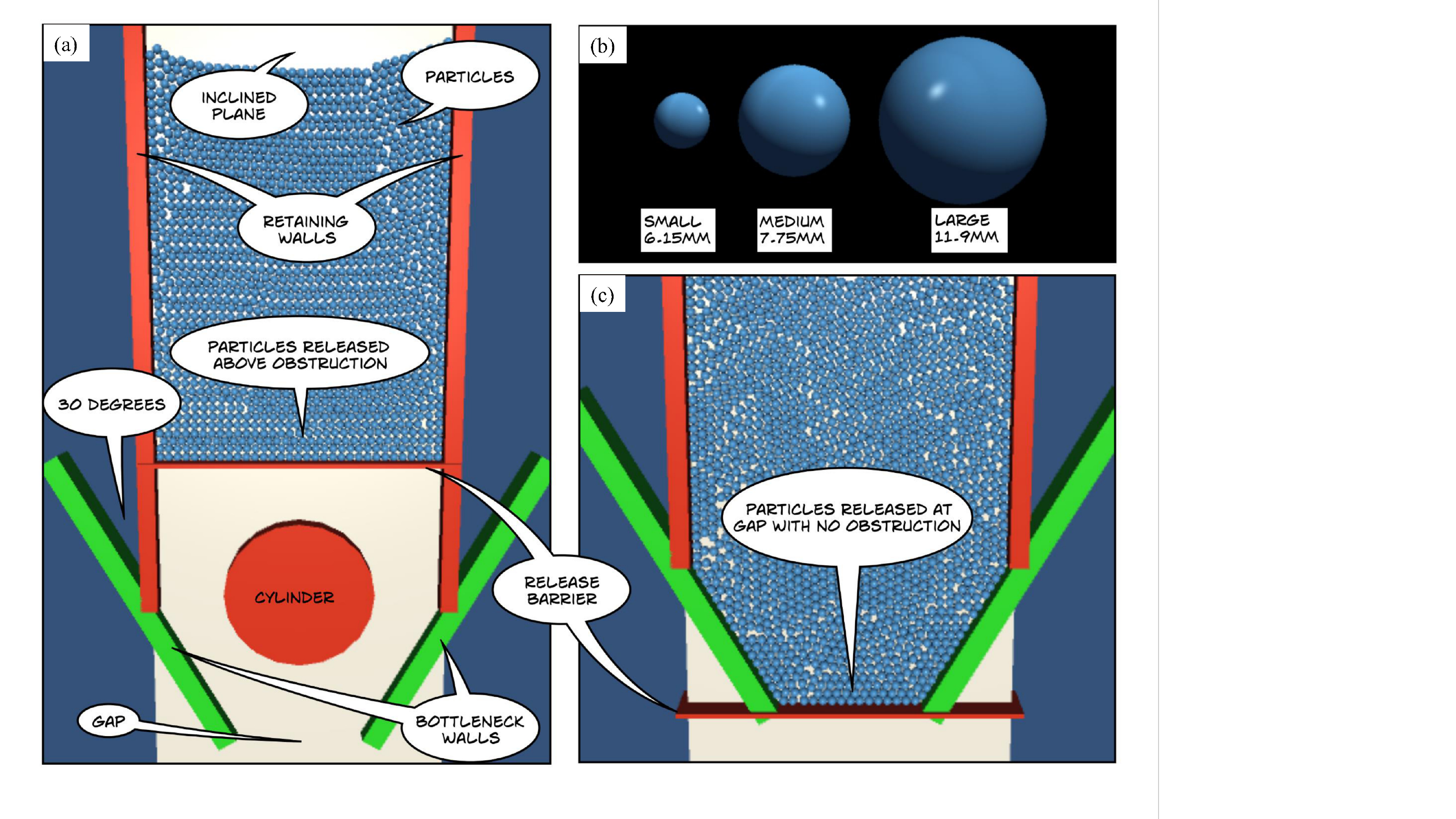} 
\caption{(a) Basic set up of the experiment with an obstruction. (b) Particles
are spheres of three different diameters. (c) Experiment of flow rate
without obstructions, the particles are released at the gap.}
\label{The Particles} 
\end{figure}

The basic set up of the experiment with obstructions is shown in Fig.
\ref{The Particles}(a), where particles are released from above
the obstruction. The particles themselves are spheres with diameters:
$6.15$ mm, $7.75$ mm and $11.9$ mm, referred to as `small', `medium',
and `large' respectively, and are shown in Fig. \ref{The Particles} 
(b). Note that without obstructions, the particles are released at
the gap as shown in Fig. \ref{The Particles}(c).

We use an inclined plane that makes an angle of $8^{o}$ with the
horizontal. The plane has retaining walls, walls that form a bottleneck
and release barriers as shown in Fig. \ref{The Particles}(a). We
pack a known number of spheres of a specified size, shown without texture
in Fig. \ref{The Particles}(b), onto the plane and release them
from rest so that they collide with the bottleneck walls, and flow
through the gap. The walls of the bottleneck make an angle of $30^{o}$
to the retaining walls.

Under gravity, the spheres move down the inclined plane, come into
contact with the bottleneck walls and flow through the gap. The number
of particles is known and the time taken for all particles to flow
through the gap is measured, from which the flow rate, $J$, is determined.
The flow rate is thus defined as the number of particles passing
through the gap per second.

The experiment is repeated for various gap widths, with the corresponding
flow rates measured. The entire experiment is performed individually
for small, medium, and large spheres. All dimensions (particles, plane,
bottleneck angles, etc) are known, as are the coefficients of friction
and restitution of the particles. As per the experiment, in our model
we release the particles at rest from the gap as shown in Fig. \ref{The Particles}
(c).

\begin{figure}[htb]
\centering
\includegraphics[width=0.95\columnwidth,keepaspectratio=true,trim=0cm 0 0cm 0,clip]{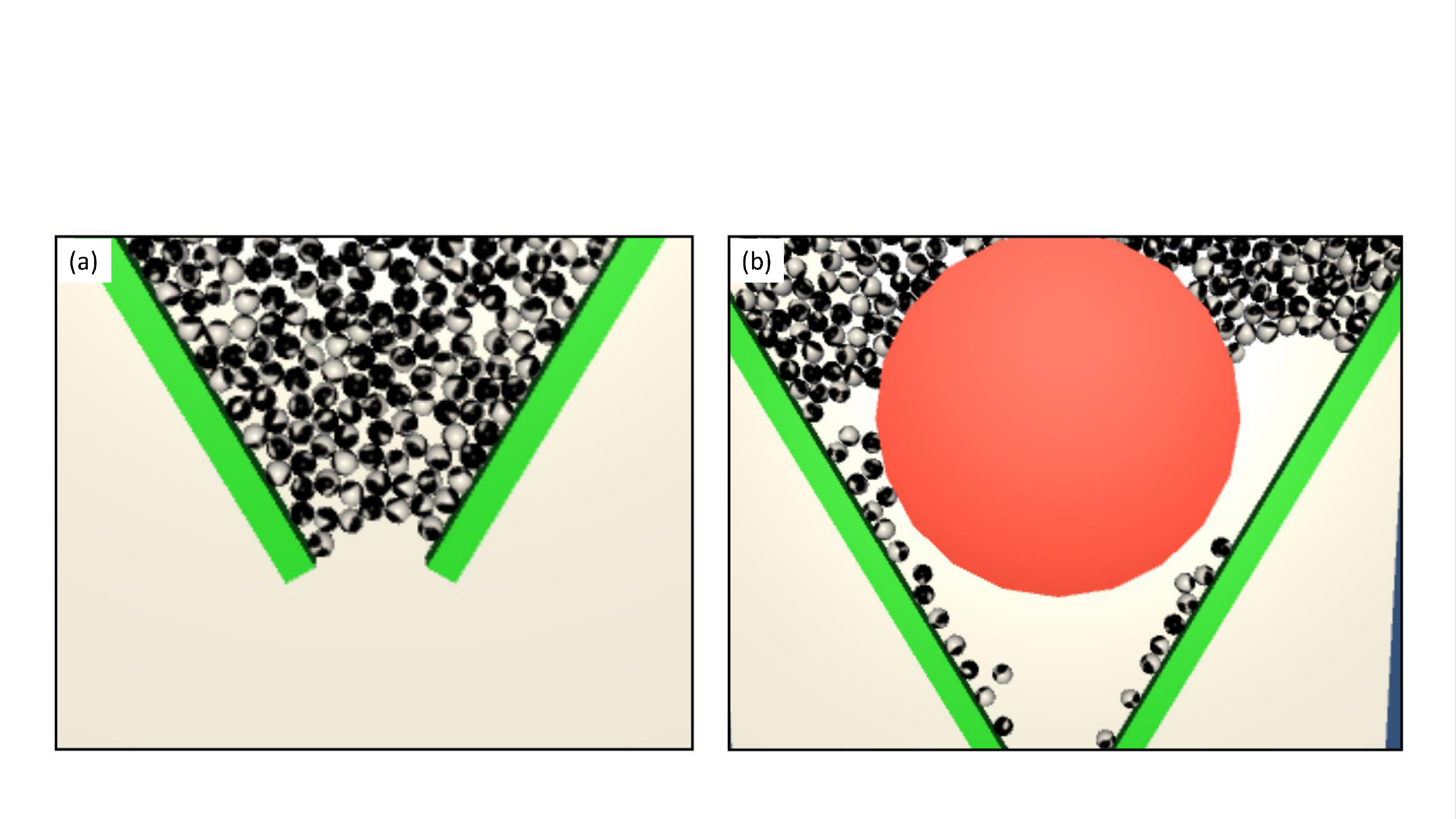}
\caption{The model exhibits clogging (a) at the gap and  between the obstruction
and the bottleneck walls (b), consistent with the experiment.}
\label{Clogs}
\end{figure}

It was found that there is a relatively sharp rise in flow rate as
the gap is increased. It is precisely this range of distances where
the flow rate transitions from clogging to intermittent to continuous
flow rates. The model also exhibits clogging as shown in Fig. \ref{Clogs} 
(a), and intermittent flow. Therefore, in order to get reliable and
reproducible flow rates, we decided to perturb the particles just sufficiently
to prevent extended periods of clogging which thereby affect flow rates.
As described by Garcimartin et al. \cite{Garcimartin}, the problem
of clogging can effectively be eliminated by applying external vibrations.

\begin{figure}[htb]
\centering
\includegraphics[width=0.95\columnwidth,keepaspectratio=true,trim=3cm 0 3cm 0,clip]{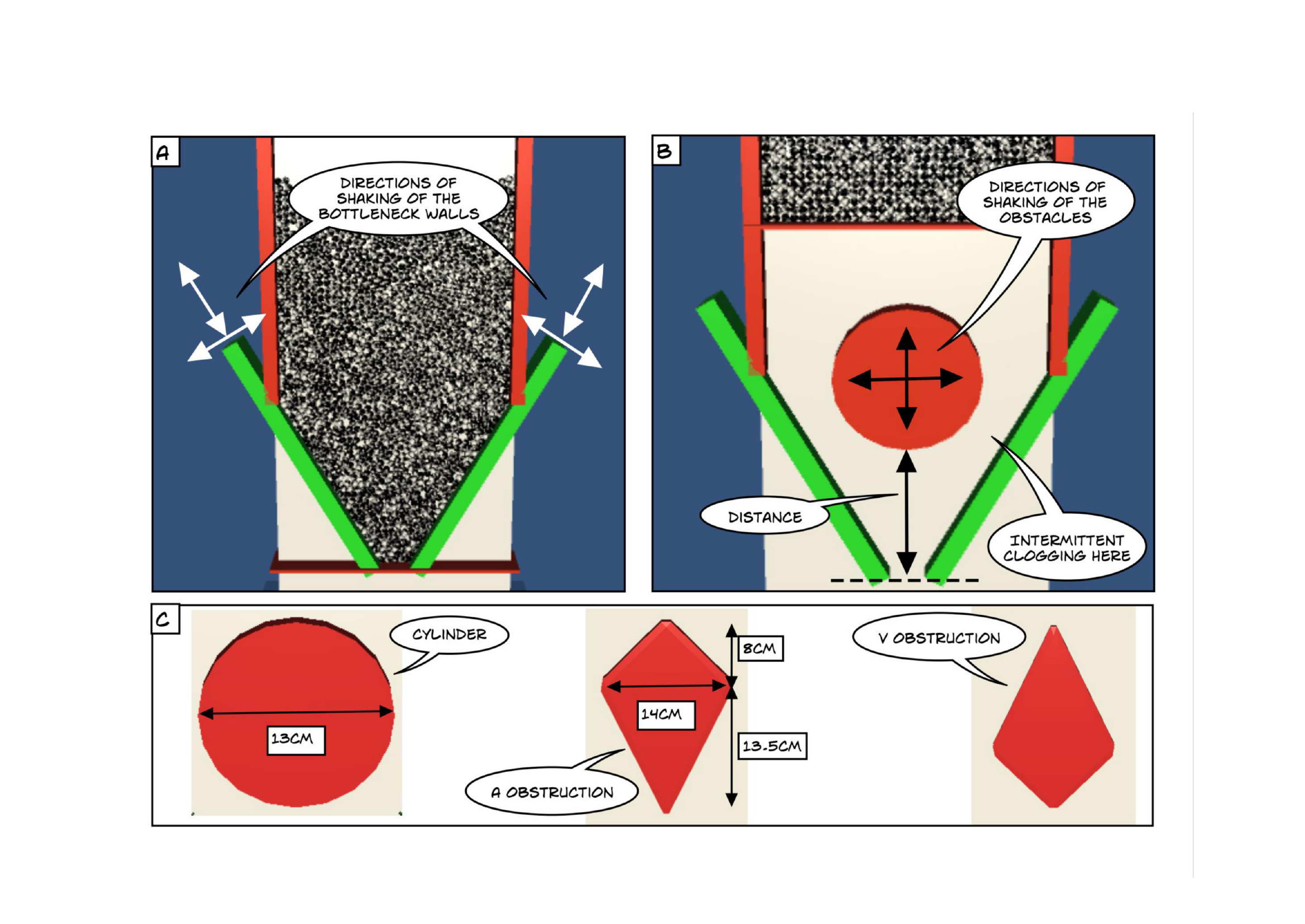}
\caption{(a) directions of shaking of the bottleneck walls in the experiment
of flow rate without obstructions. (b) directions of shaking in the
experiment of flow rate with obstructions. Shaking is necessary to
avoid clogging between the obstruction and the bottleneck walls. (c)
obstructions used in experiments. The V obstruction is the A obstruction
rotated through $180^{o}$.}
\label{Shaking} 
\end{figure}

In a similar way, we use the technique of `shaking'. In the case
of flow rate without obstructions, the bottleneck walls were shaken
as shown in Fig. \ref{Shaking}(a) from a maximum of $0.5$ mm diminishing
to zero over one second, every five seconds. The parameters that define
a shake are: shake amplitude, direction of shake, shake duration and
time between shakes. These parameters can be set arbitrarily in the
model and, whilst other times and distances were tested, the above
were found to be just sufficient to yield reproducible results.

We varied the values of coefficient of friction and coefficient of
restitution in the model until we got good agreement with experimental
results. It was found that varying the coefficient of restitution
had little to no effect on flow rate.

\begin{figure*}[htb]
\centering
\includegraphics[width=0.85\textwidth,keepaspectratio=true]{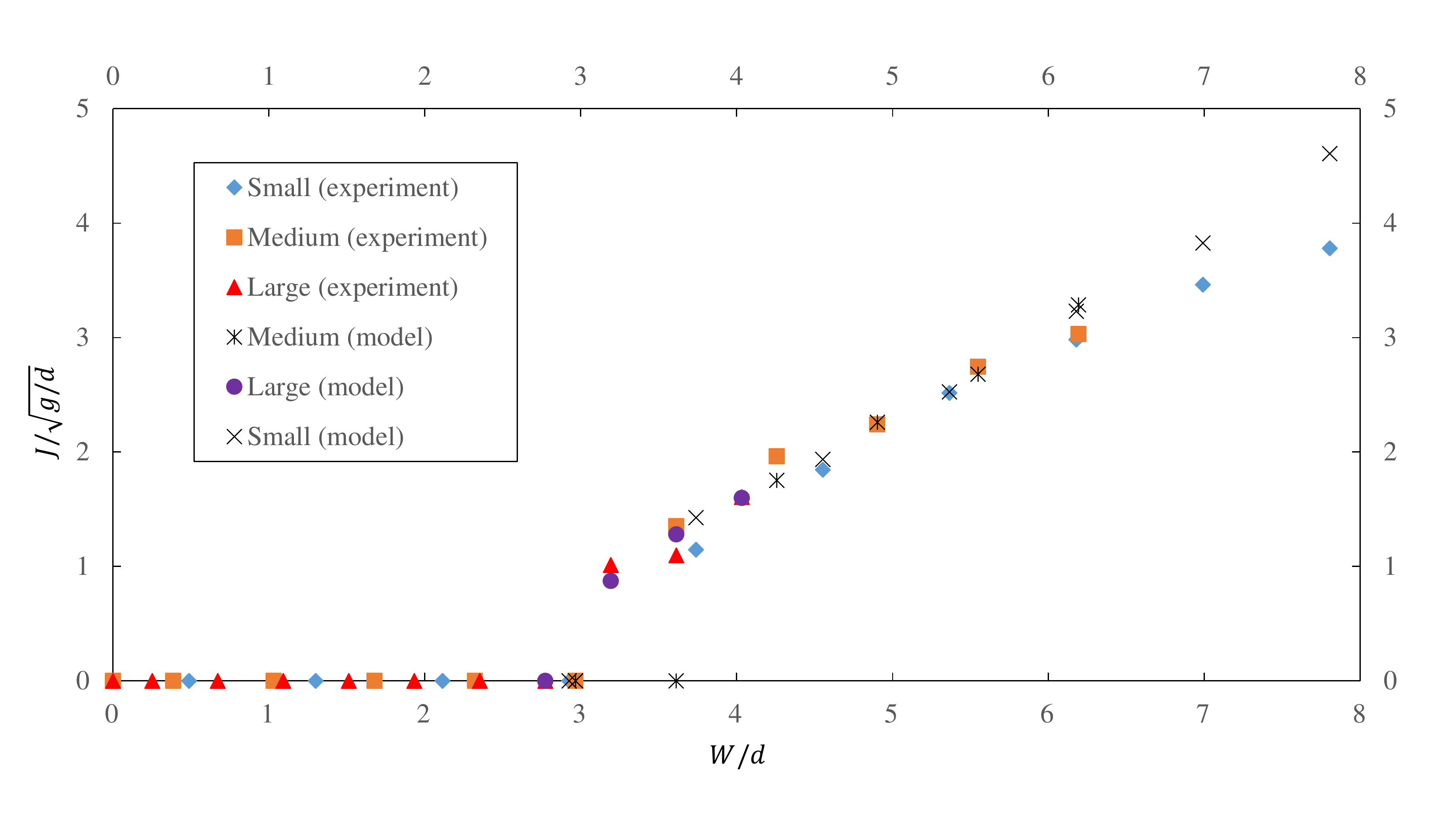} 
\caption{Flow rate without obstruction versus gap in dimensionless form. Comparison
of experiment with simulation using spheres of three different diameters.
$J$ is the flow rate, $g$ is the acceleration due to gravity, $d$
is the particle diameter and $W$ is the gap width.}
\label{Calibration} 
\end{figure*}

We found a coefficient of friction of $0.24$ yielded model results
that gave good agreement with experimental results. The experimentally
measured value of the spheres was found to be $0.28\pm0.4$. The results
for each diameter sphere are shown in Fig. \ref{Calibration}. The
model results agree with the experimental results quite well.

\section{Flow rate with obstructions}

\label{sec:With Obstructions}

The experiment is repeated with medium spheres (of diameter $7.75$
mm). However, this time, obstructions (shown in Fig. \ref{Shaking} 
(c)) are placed at varying distances away from the gap, which is now
fixed at a width of $3.3$ cm.

It was found that there was a relatively sharp rise in flow rate as
an obstruction is moved away from the gap. It is precisely this range
of distances where the flow rate transitions from clogging to intermittent
to continuous flow rates. The model also exhibited clogging as shown
in Fig. \ref{Clogs}(b), and intermittent flow. Therefore, in order
to get reliable and reproducible flow rates, we decided to perturb
the particles just sufficiently to prevent extended periods of clogging
which thereby affect flow rates.

\begin{figure}[htb]
\centering
\includegraphics[width=0.48\textwidth,keepaspectratio=true,trim=1cm 0 1cm 6cm,clip]{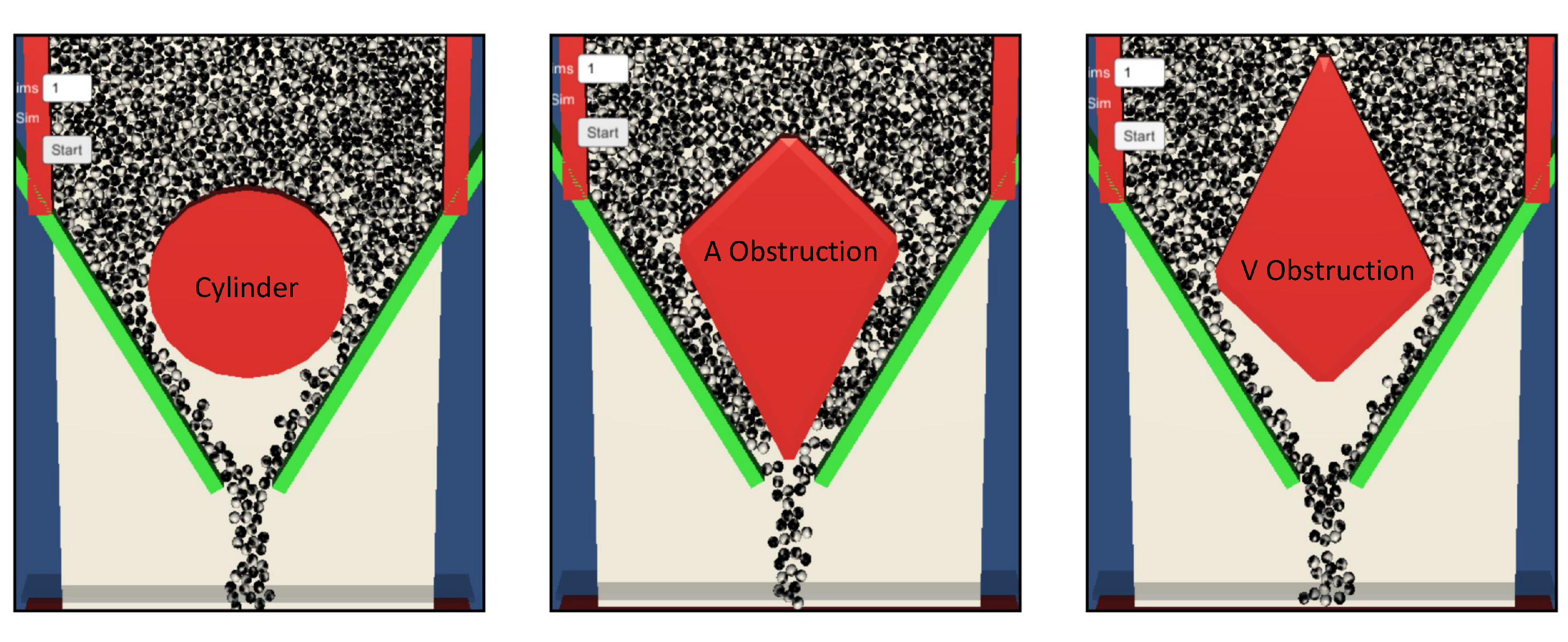}
\caption{Peak flow rates with obstructions, where the `waiting room' effect
is present for the cylinder and the V obstruction but not for the A obstruction.}
\label{Peak Flowrates} 
\end{figure}

With the gap set at $3.3$ cm, no clogging was observed for spheres
of $7.75$ mm at the gap. However, with the presence of an obstruction,
clogging was observed in areas between the obstruction and the bottleneck
walls as shown in Fig. \ref{Clogs}(b) and Fig. \ref{Shaking}(b).
To get consistently measurable flow rates in the model, once again,
we decided to perturb the system using the technique of `shaking',
described above. This time, we decided to shake the obstruction also
shown in Fig. \ref{Shaking}(b). The obstruction was shaken by a
maximum of $1$ mm diminishing to zero over one second, every five
seconds. The freely specifiable parameters that define a `shake' are
the same as those for shaking the bottleneck walls. Figure \ref{Peak Flowrates}
shows the maximum flow rates for the three obstructions, where the
`waiting room' effect, discussed in Ref. \cite{Alonso-Marroquin}, occurs
for the cylinder and the V obstruction but not the A obstruction.

\begin{figure*}[htb]
\centering
\includegraphics[width=0.90\textwidth,keepaspectratio=true]{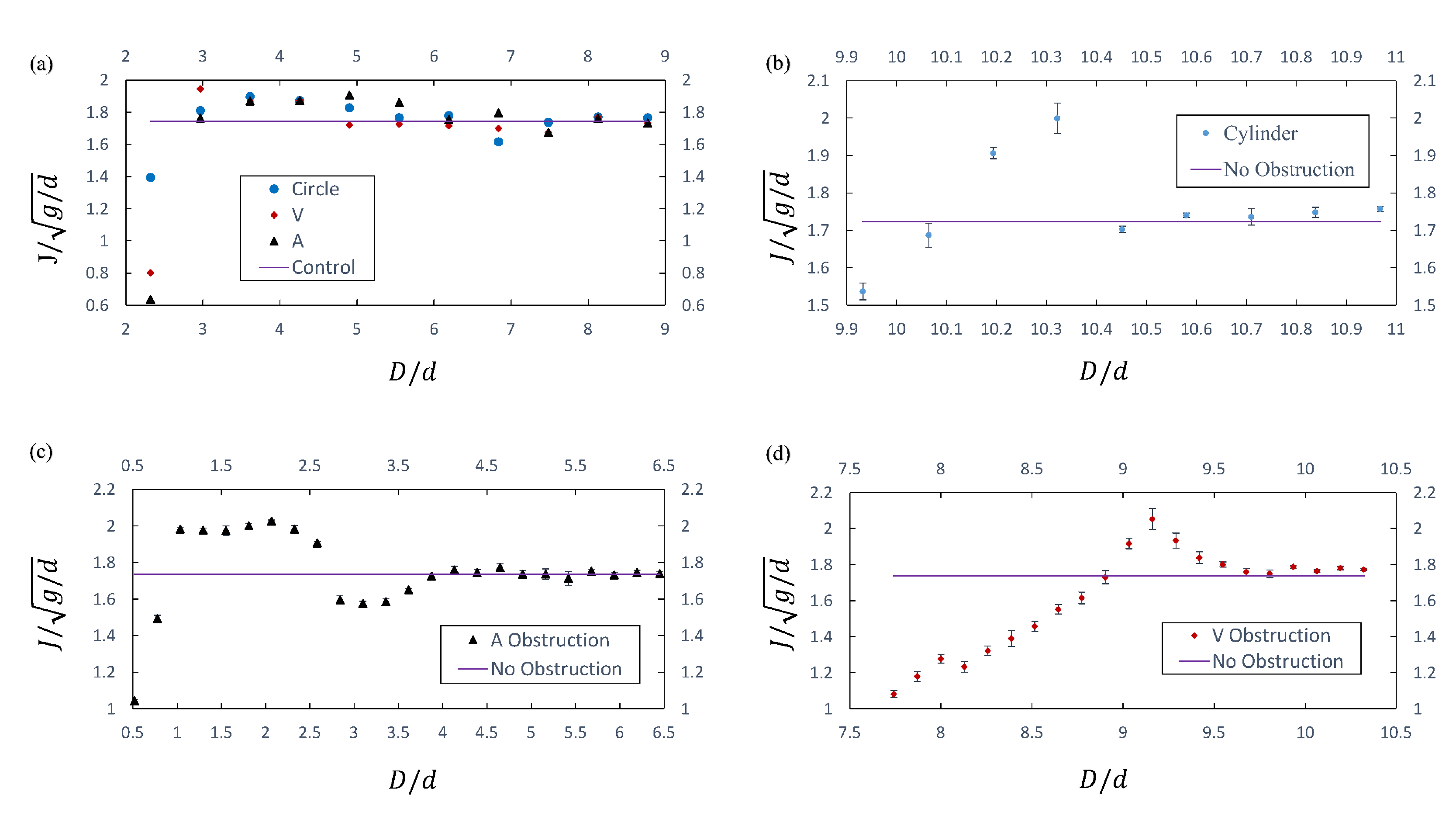}
\caption{Results of flow rate with obstructions versus distance in dimensionless
form. $J$ is the flow rate, $g$ is the acceleration due to gravity,
$d$ is the particle diameter, and $D$ is the distance from the obstruction
to the gap. The experimental results in (a) show that each obstruction
has a range of distances where flow rate is greater than with no obstruction.
The numerical results (b-d) show that the model also predicts that
each obstruction has a range of distances from the gap where the flow rate
is greater than flow rate with no obstruction.} 
\label{Obstructions} 
\end{figure*}

Figure \ref{Obstructions} shows flow rates as measured experimentally
and with the model. Both sets of results indicate increases in flow rates
for certain ranges of distance of obstruction to the gap. For the
model, each graph shows error bars and data that were more than two
standard deviations away from the mean were not included in the statistical
analysis. This was necessary as, even with the shaking technique described
above, the particles did occasionally clog for extended periods of
time, thereby affecting the flow rate measurements.

Clearly, the model shows all obstructions improve flow rate for a range
of distances, and reassuringly, the further the obstruction is moved
away from the gap, the closer we get to flow rates without obstruction.

In the case of the cylinder, due to its shape and size relative to
the bottleneck, we could only start at distances $\sim7$ cm from
the gap. We see that there is a steady rise in flow rate as the cylinder
is moved further away from the gap. A maximum flow rate is reached
and then a slightly sharper decrease is reached until we get flow rates
the same as if no obstruction is present.

\textcolor{black}{In the case of the A obstruction, due to its shape
and size relative to the bottleneck, we were able to place the obstruction
very close to the gap, which is why distances start at $\sim0.4$
cm from the gap. We see a sharp rise in flow rate as the obstruction
is moved away from the gap, and then we see a leveling out of the
flow rate over a range of distances from $0.8$ cm to $2.8$ cm. There
is a small peak in this range, but we consider this a statistical
fluctuation.}

\textcolor{black}{Moving further away, from $2.8$ cm to $3.2$ cm,
we see a sharp decrease in flow rate to values that are actually below
flow rates with no obstacle. This is an intriguing phenomenon, showing
what we might expect: obstructions decrease flow rates.}

The flow rate remains constant to a distance of $3.6$ cm and then
starts rising. At a distance of $4$ cm, the flow rate is approximately
equal to that with no obstructions.

In the case of the V obstruction, due to its shape and size relative
to the bottleneck, we could only start at distances $\sim6$ cm from
the gap. We see that there is a steady rise in flow rate that is slower
than that of the A obstruction,\textcolor{black}{{} as the V obstruction
is moved further away from the gap.} A maximum flow rate is reached,
higher than that with no obstruction, and then a decrease occurs until
we get flow rates the same as if no obstruction were present.

\section{Limitations of PhysX engine}

\label{sec:Limitations}

We observed that the model exhibits continuous flow, intermittent
flow, and clogging, all phenomena that have been identified and observed
experimentally. The model was quite sensitive to the absolute sizes
of the geometric structures used. This is true for all physics engines,
as they have to be `tuned' to the range we would like them to work
most accurately in. For example, an engine might be tuned to work
best with objects whose sizes are of the order of meters but it will
not work for objects at the scale of millimeters. To overcome this
limitation, we ran the simulations in the dimensions where the performance
of the engine was optimal, and then used dimensional analysis to extrapolate
the result to the real experimental conditions.

We found that the particles in the model needed to be perturbed more
often than in actual experiments, in order to get consistent flow rates.
This can be attributed to a real time optimization employed by PhysX,
known as `sleeping'. When a particle's angular and/or linear velocity
falls below certain threshold values for more than a few frames, these
velocities are set to zero and the particle goes into a `sleep' state,
in which no collision detection occurs and hence the particle's velocity
remains at zero. The particle `wakes up' when it is subjected to net
forces. The reason for this optimization is that it relieves the processor
of having to perform needless computations especially with regard
to collision detection, thereby allowing much larger numbers of particles
to be present and only performing the necessary computations as required.
For this study, we left the threshold `sleep velocities' at their
default values.

In spite of these limitations, we can confirm that, as reported in
various experiments, it is possible to increase flow rate of discrete
particles by placing an obstruction at a suitable location near a
gap.

\section{Conclusions}

\label{sec:Conclusion}

We have developed a model that exhibited reasonable quantitative accuracy
in the case of no obstructions, and good qualitative agreement when
obstructions were introduced. As in experiments, we showed that the
flow exhibits clogging, intermittent flow rates, and continuous flow rates.
The model also confirms experimental results that placement of an
obstruction near a gap can actually increase flow rate through the
gap. Better experimental agreement with obstructions will be possible
if the roundness of the obstructions is more accurately modeled.

The model is quite flexible in that many parameters can be changed
via user input, and without modifying the actual code. In this way,
the model can be used as a convenient tool to suggest further experiments
and allows us to investigate different obstruction shapes, both of
which may help us gain a greater understanding of granular flow.

We have also described a suite of freely available realtime 3D software,
which is mature and has many online resources in the form of documentation
and tutorials and is being actively developed, which can be used
together to create simulations of interest in the area of granular
flow.

\begin{acknowledgements}
We thank Ivan Gojkovic and Jaeris Wu for performing the experimental part of this paper.
\end{acknowledgements}

\end{document}